\documentclass[twocolumn,prl,showpacs,floatfix]{revtex4}

\usepackage{subfigure}
\usepackage{graphicx}

\begin{document}

\preprint{DRAFT}
\date{September 6, 2011}

\title{Superconducting Radio Frequency Cavities
as Axion Dark Matter Detectors}

\author{P.~Sikivie}

\affiliation{Department of Physics, University of Florida, 
Gainesville, FL 32611, USA}

\begin{abstract}

A modification of the cavity technique for axion dark matter 
detection is described in which the cavity is driven with input 
power instead of being permeated by a static magnetic field.  
A small fraction of the input power is pumped by the axion 
field to a receiving mode of frequency $\omega_1$ when the 
resonance condition $\omega_1 = \omega_0 \pm m_a$ is satisfied, 
where $\omega_0$ is the frequency of the input mode and $m_a$ 
the axion mass.  The relevant form factor is calculated for 
any pair of input and output modes in a cylindrical cavity.  
The overall search strategy is discussed and the technical 
challenges to be overcome by an actual experiment are listed.

\end{abstract}

\maketitle

Shortly after the Standard Model of elementary particles was established, 
the axion was postulated \cite{axion} to explain why the strong interactions 
conserve the discrete symmetries P and CP.  Further motivation for the 
existence of such a particle came from the realization that cold axions 
are abundantly produced during the QCD phase transition in the early 
universe and that they may constitute the dark matter \cite{axdm}.  
Moreover, it has been claimed recently that axions are the dark matter 
\cite{CABEC} because cold axions predict a specific phase space distribution 
for the halos of isolated disk galaxies which is supported by observation.  
These considerations provide strong motivation for developing new approaches
to axion dark matter detection.  

Axion properties depend mainly on a single parameter $f_a$, called the axion 
decay constant.  In particular the axion mass ($\hbar = c = 1$)
\begin{equation}
m_a \simeq 6 \cdot 10^{-6}~{\rm eV}~{10^{12}~{\rm GeV} \over f_a}
\label{mass}
\end{equation} 
and its coupling to two photons:
\begin{equation}
{\cal L}_{a\gamma\gamma} = 
g_\gamma {\alpha \over \pi} {a(x) \over f_a} \vec{E}(x)\cdot\vec{B}(x)~~~\ .
\label{emcoupl}
\end{equation}
Here $a(x)$ is the axion field, $\vec{E}(x)$ and $\vec{B}(x)$ the electric 
and magnetic fields, $\alpha$ the fine structure constant, and $g_\gamma$
a model-dependent coefficient of order one. Although $f_a$ was first thought 
to be of order the electroweak scale, its value is in fact arbitrary a-priori
\cite{invis}.  However, the limits from unsuccessful axion searches in particle 
and nuclear physics experiments combined with constraints from stellar evolution 
imply $f_a \gtrsim 3 \cdot 10^9$ GeV \cite{axrev}.  Furthermore, the requirement 
that cold axions do not overclose the universe implies $f_a \lesssim 10^{12}$ GeV 
\cite{axdm}.  The constraints leave open an `axion window': 
$2 \cdot 10^{-3} \lesssim m_a \lesssim 6 \cdot 10^{-6}$ eV.

It was pointed out in ref. \cite{axdet} that dark matter axions may be detected 
on Earth through their conversion to microwave photons in a cavity permeated by 
a strong magnetic field.  Using this principle and building on earlier efforts
\cite{early}, the ADMX experiment has reached sufficient sensitivity to detect 
halo axions at the expected density \cite{ADMX,HiRes}.  When SQUIDs are installed 
as front-end amplifiers \cite{SQUID} and its cavity is cooled to ~50 mK, ADMX will 
be able to detect, or set limits on, halo axions at even a fraction of the halo 
density.  The present detector would find halo axions immediately (in a minute 
or so) if the axion mass were known.  Unfortunately the axion mass is poorly 
constrained.  The value of the axion mass for which axions provide the observed 
density of cold dark matter ($\Omega_a = \Omega_{\rm CDM} = 0.23$) is only known 
to be of order $10^{-5}$ eV with large uncertainties, at least a factor 10 in 
either direction.

The cavity technique is most convenient in the 500 MHz to 1 GHz range, 
corresponding to 2.068 $< m_a < 4.136~\mu$eV, because those are typical 
resonant frequencies for the lowest TM mode of cavities large enough 
(fraction of one (meter)$^3$) to produce a signal, given that available 
(large volume) magnetic field strengths are of order 8 T.  Only the lowest 
TM mode of a cylindrical cavity with axis in the direction of the magnetic 
field has a sizeable coupling to halo axions.  To extend the search to higher 
frequencies one must fill the available magnetic field volume with an increasing 
number of smaller cavities, kept in tune and in phase during the search.  Although 
the principle of this method has been demonstrated \cite{Hagmann}, it presents a 
challenging engineering problem in practice.

A possible variation on the cavity permeated by a static magnetic field is 
a cavity driven with power $P_0$ at the frequency $\omega_0$ of one of the 
cavity modes, hereafter called the input mode \cite{Mel}.  Let $\omega_1$ 
be the frequency of another cavity mode, called the signal mode.  When the 
resonance condition $\omega_1 = \omega_0 \pm E_a$ is satisfied, where $E_a$ 
is the energy of halo axions, the steady state power $P_1$ in the signal 
mode through the coupling of Eq.~(\ref{emcoupl}) is 
\begin{equation}
{P_1 \over P_0} = {1 \over 2} g^2 \rho_a C {Q_0 Q_1 \over \omega_0^2}
\label{power}
\end{equation}
where $g \equiv {g_\gamma \alpha \over \pi f_a}$, $Q_0$ and $Q_1$ are 
the quality factors of the cavity in the input and signal modes, and 
$\rho_a$ is the fraction of the local axion dark matter density with 
energy dispersion $\delta E_a < \delta\omega_1 \equiv \omega_1/Q_1$.  
The input power is assumed to be monochromatic. The dimensionless 
form factor $C$ is 
\begin{equation} 
C = {\omega_0 \over \omega_1} 
{\left(~{1 \over \epsilon}~
\int_V d^3 x \vec{B}_0(\vec{x})\cdot\vec{E}_1(\vec{x})\right)^2 
\over \left(\int_V d^3 x \vec{E}_0(\vec{x})\cdot\vec{E}_0(\vec{x})\right)
\left(\int_V d^3 x \vec{E}_1(\vec{x})\cdot\vec{E}_1(\vec{x})\right)}~~~\ ,
\label{form}
\end{equation}
where $\epsilon$ is the dielectic constant inside the cavity.  
$C$ is invariant under the interchange of input and signal modes.  
Eq.~(\ref{power}) implies therefore that, all other things being 
equal, it is best to drive the cavity in the lowest frequency mode 
of any given pair.  

Let $v$ and $\delta v$ be respectively the velocity and velocity 
dispersion of the axions.  Eq.~(\ref{power}) assumes that the de 
Broglie wavelength $(m_a v)^{-1}$ is large compared to the cavity's 
linear dimensions.  Furthermore, the energy dispersion must be less 
than the bandwidth of the signal mode.  The axion energy is the sum 
of rest mass and kinetic energy: $E_a = m_a (1 + {1 \over 2} v^2)$.  
Hence $\delta E_a \simeq m_a v \delta v$. It is often assumed 
that the velocity distribution of halo axions is Maxwellian 
with velocity dispersion $\delta v \simeq 300$ km/s, in which case 
$\delta E_a \simeq 10^{-6} m_a$.  However an isolated disk galaxy such 
as the Milky Way continually accretes the dark matter surrounding it 
and the resulting flows do not thermalize over the age of the universe 
\cite{Ips}.  The infall flows produce peaks in the velocity spectrum 
of dark matter at any given spatial location.  Furthermore, the flows 
form caustics.  Caustics are surfaces in physical space where the dark 
matter density is very large.  Evidence was found for caustic rings in 
the Milky Way and in other isolated disk galaxies.  The evidence is 
summarized in ref. \cite{MWhalo}.  If an observer in a galactic halo 
is close to a caustic, his local dark matter velocity distribution is
dominated by a single flow or a single pair of flows.  The evidence for 
caustic rings in the Milky Way implies that we are in fact located close 
to a caustic and that our local halo density is dominated by a single 
flow, dubbed the `Big Flow', with known velocity vector, density of 
order 1 GeV/cc, and velocity dispersion less than 50 m/s \cite{BF}.  
The energy dispersion of the Big Flow is thus less than approximately 
$10^{-10} m_a$.

Superconducting radio frequency cavities have quality factors as large 
as $10^{10}$ \cite{Padamsee}.  Consider however that, even if there are 
axion flows with negligible velocity dispersion, increasing the quality 
factor indefinitely does not generally yield the best search strategy 
because high $Q$ implies a long time to excite the signal mode.  The 
time necessary for the power in the signal mode to rise to 90\% of 
the steady state power $P_1$ is $t_Q \simeq {Q_1 \over \nu_1}$ where 
$\nu_1 = {\omega_1 \over 2 \pi}$. The total time spent at a given 
cavity tune is $t = t_Q + t_d$ where $t_d$ is the measurement 
integration time.  The signal to noise ratio of the search is 
given by the radiometer equation \cite{Dicke}
\begin{equation}
s/n = {P_1 \over T_{\rm n}} \sqrt{t_d \over B}~~~\ ,
\label{radio}
\end{equation}
where $T_{\rm n}$ is the total noise temperature (thermal noise from 
the cavity plus electronic noise from the receiver chain) and $B$ is 
the bandwidth.  In searching for cold flows, the best strategy is to 
make the bandwidth as narrow as possible, i.e. $B = 1/t_d$.  Eq.~(\ref{radio})
implies then that $t_d \propto 1/P_1 \propto 1/Q_1$.  Because $t = t_Q + t_d$
with $t_Q \propto Q_1$ and $t_d \propto Q_1^{-1}$, the optimal $Q_1$,
the one that maximizes $s/n$ for given $t$, is such that $t_Q = t_d$ and 
hence $t = {2 Q_1 \over \nu_1}$.  Therefore, to allow a search over a 
factor two in axion mass per year, $Q_1$ should not be larger than 
approximately  $10^8 \left({\nu_1 \over {\rm GHz}}\right)^{1 \over 2}$.  

When $B = 1/t_d$, signal averaging is not possible and the noise 
is exponentially distributed, as in the HiRes channel of the ADMX 
experiment \cite{HiRes}.  To have less than 100 false positives due 
to noise fluctuations per factor of two in frequency range covered, 
the probability that such a fluctuation exceed the signal must be 
less than approximately $10^2/Q_1$ per cavity bandwidth.  For 
$Q_1 \sim 10^8$ and exponentially distributed noise, this requires  
$s/n \simeq 15$.  Eq.~(\ref{radio}) with $B = t_d^{-1} = \nu_1/Q_1$ 
implies the minimum detectable signal power 
\begin{equation}
P_1 = 2.1 \cdot 10^{-20}~{\rm W} \left({T_n \over 10~{\rm K}}\right)
\left({s/n \over 15}\right) \left({\nu_1 \over {\rm GHz}}\right)
\left({10^8 \over Q_1}\right)~~~\ .
\label{sigpow}
\end{equation}
To determine how much input power $P_0$ will be needed, it is useful 
to write
\begin{equation}
{g^2 \rho_a \over m_a^2} = 1.5 \cdot 10^{-43} 
\left({g_\gamma \over 0.36}\right)^2
\left({\rho_a \over {\rm GeV/cc}}\right)~~~\ .
\label{small}
\end{equation}
$g_\gamma = 0.36$ in the DFSZ axion model, whereas $g_\gamma = -0.97$
in the KSVZ model.   Combining Eqs.~(\ref{power}), (\ref{sigpow}) and 
(\ref{small}), one obtains
\begin{equation}
P_0 = 2.8 \cdot 10^7~{\rm W} 
\left({\omega_0 \over m_a}\right)^2 
\left({\nu_1 \over {\rm GHz}}\right){1 \over C}
\left({10^8 \over Q_1}\right)^2 \left({10^8 \over Q_0}\right)
\label{much}
\end{equation}
for the benchmark values of $T_{\rm n}$, $s/n$, $g_\gamma$ and $\rho_a$ 
used above.  A large fraction of the power $P_0$ can be recycled and 
need not contribute to the energy cost.  Indeed the power loss in the 
cavity is the sum of losses in the walls and losses through the hole 
through which the signal is coupled out of the cavity: 
$Q_0^{-1} = Q_w^{-1} + Q_h^{-1}$.  If for example $Q_w = 10^{10}$ and 
$Q_h = 10^8$, the power dissipated in the cavity walls is only one 
hundred of the input power circulating through the cavity.  The 
remaining 99\% may be recycled.  

Consider a cylindrical cavity, i.e. with constant cross-section relative to an 
axis $\hat{z}$.  The cavity modes are labeled as usual as TE$_{mnp}$, TM$_{mnp}$ 
and TEM$_p$ \cite{Jackson} where $p$ refers to the $z$-dependence of the mode.  
Using Eq.~(\ref{form}), one finds that $C = 0$ unless either the input or the 
signal mode, or both, is TE.  There are therefore three cases in which the 
form factor does not vanish: TEM $\leftrightarrow$ TE, TM $\leftrightarrow$ TE, 
and TE $\leftrightarrow$ TE.

For TEM$_p \leftrightarrow$ TE$_{mnp^\prime}$, the form factor is
\begin{equation}
C = {4 \over \pi^2} \left({2 p^\prime \over p^{\prime 2} - p^2}\right)^2
{\omega \over \epsilon^2 \omega^\prime} \gamma^2 
{\left(\int_S d^2 x~\phi\varphi \right)^2 \over 
\left(\int_S d^2 x~\phi^2 \right) 
\left(\int_S d^2 x~(\vec{\nabla}\varphi)^2\right)}
\label{TEMTE}
\end{equation}
if $p + p^\prime$ is odd.  $\varphi$ defines the TEM mode through 
$\vec{E} = - \vec{\nabla} \varphi(x,y) \sin\left({\pi p z \over L}\right)$
and $\phi$ defines the TE mode through
$\vec{E} = - {i \omega \over \gamma^2} \hat{z}\times\vec{\nabla}\phi(x,y)
\sin\left({\pi p^\prime z \over L}\right)$. $L$ is the length of the cavity,
$S$ is its cross-sectional area, 
$\omega = {\pi p \over \sqrt{\epsilon} L}$ is the frequency of the TEM 
mode and $\omega^\prime = \sqrt{{1 \over \epsilon}
\left[\gamma^2 + \left({\pi p^\prime \over L}\right)^2\right]}$
that of the TE mode.

For TM$_{mnp} \leftrightarrow$ TE$_{m^\prime n^\prime p^\prime}$, 
the form factor is
\begin{equation}
C = {4 \over \pi^2} \left({2 p^\prime \over p^{\prime 2} - p^2}\right)^2
{\gamma^{\prime 2} \omega \over \gamma^2 \omega^\prime}
{\left(\int_S d^2 x~\phi\psi \right)^2 \over
\left(\int_S d^2 x~\phi^2 \right)
\left(\int_S d^2 x~\psi^2\right)}
\label{TMTE}
\end{equation}
if $p + p^\prime$ is odd.  $\psi$ defines the TM mode through
$\vec{B} = {i \epsilon \omega \over \gamma^2} \hat{z} \times 
\vec{\nabla} \psi(x,y)~\cos\left({\pi p z \over L}\right)$.
$\omega = \sqrt{{1 \over \epsilon}
\left[\gamma^2 + \left({\pi p \over L}\right)^2\right]}$ 
and $\omega^\prime = \sqrt{{1 \over \epsilon}\left[\gamma^{\prime 2} + 
\left({\pi p^\prime \over L}\right)^2\right]}$.  

Finally, for TE$_{mnp} \leftrightarrow$ TE$_{m^\prime n^\prime p^\prime}$
\begin{equation}
C = {1 \over \epsilon^2 L^2 \omega \omega^\prime \gamma^2 \gamma^{\prime 2}}
\left({4 p p^\prime \over p^{\prime 2} - p^2}\right)^2
{\left(\int_S d^2 x 
(\vec{\nabla}\phi \times \vec{\nabla}\phi^\prime)\cdot \hat{z}\right)^2 
\over   
\left(\int_S d^2 x \phi^2 \right)
\left(\int_S d^2 x \phi^{\prime 2}\right)} 
\label{TETE}
\end{equation}
if $p + p^\prime$ is odd.  

In all three cases, $C = 0$ is $p + p^\prime$ is even.  Generally speaking, 
$C$ is largest when the numbers characterizing the input and signal modes 
differ as little as possible.  In particular, setting $p^\prime = p + 1$ 
maximizes $C$ with respect to $p^\prime$ for given $p$.

The power dissipated in the cavity walls is:
\begin{eqnarray}
P_w = 2.8 \cdot 10^5~{\rm Watt}
\left({\omega_0 \over m_a}\right)^2 {1 \over C}
\left({10^8 \over Q_1}\right)^2
\left({10^{10} \over Q_w}\right)\nonumber\\
\left({T_{\rm n} \over 10~{\rm K}}\right)
\left({\rho_a \over {\rm GeV/cc}}\right)
\left({g_\gamma \over 0.36}\right)^2
\left({s/n \over 15}\right) \left({\nu_1 \over {\rm GHz}}\right)\ .
\label{less}
\end{eqnarray}
It  must be extracted to keep the cavity cold.  The second law of 
thermodynamics implies that the (room temperature) power needed to 
do this is equal or larger than ${T_h \over T_c} P_w$ where $T_c$ is 
the cavity temperature and $T_h$ is room temperature.  In practice, 
it may be a few times that.  For example, the Linde L280/LR280 
refrigerator removes 900 W at 4.4 K while consuming 250 kW in 
electrical power.  A very large refrigerator still has only a 
cooling power of a few kW at 4.4 K.  The power dissipated in the 
cavity walls can be made of order kW by exploiting the factor 
$({\omega_0 \over m_a})^2$ on the RHS of Eq.~(\ref{less}).  To 
have $\omega_0 << m_a$ we drive the cavity in a TEM mode and 
look therefore for output power in a TE mode.  

\begin{figure}
\includegraphics[width=0.9\columnwidth]{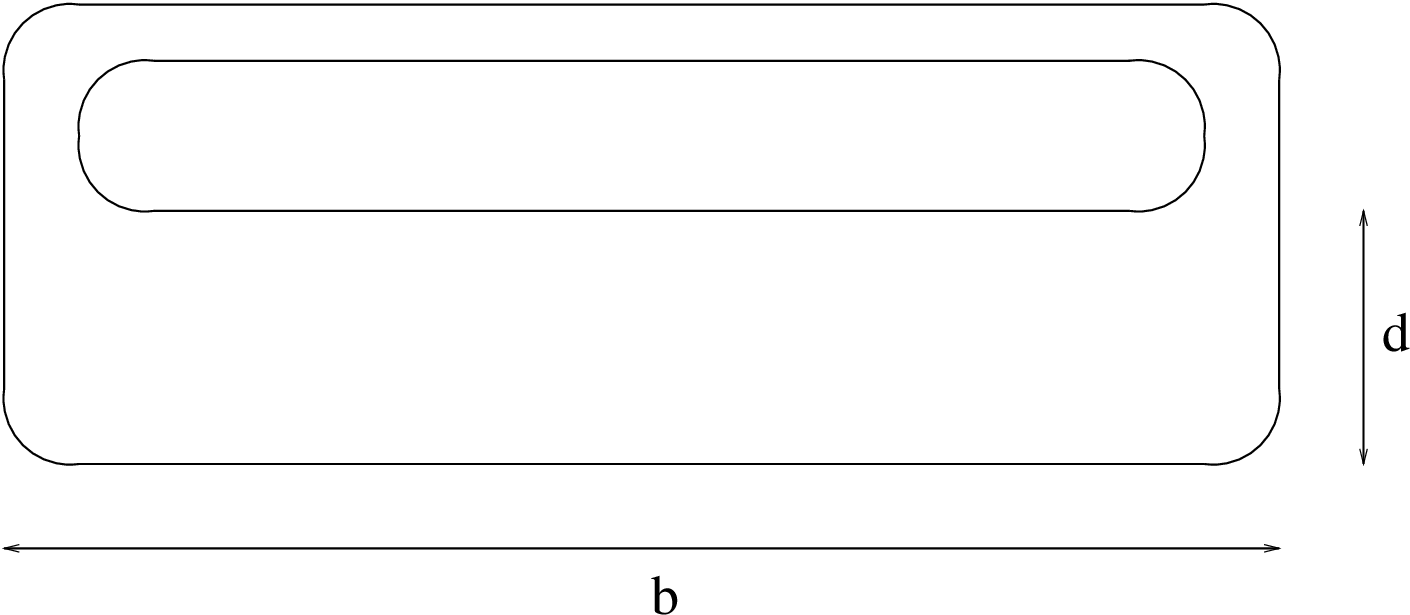}
\caption{Cross-section of a superconducting cavity to search for
axion stimulated TEM $\rightarrow$ TE transitions.}
\label{xsh}
\end{figure}

Consider a cylindrical cavity with the cross-sectional 
shape shown in Fig. \ref{xsh}. It has a single series of TEM$_p$ 
modes labeled by $p$ = 1, 2, 3 ... , and a set of TE$_{01p^\prime}$ 
modes for which $\gamma \simeq {\pi \over d}$.  In the limit $b >> d$, 
the form factor for TEM$_1 \rightarrow$ TE$_{012}$ transitions is 
\begin{equation} 
C = 0.584~{\omega_0 \over \omega_1} {1 \over \epsilon^2} 
\label{ff} 
\end{equation}
where $\omega_0 = {\pi \over L}$ is the frequency of the TEM$_1$ mode 
and $\omega_1 \simeq \pi \sqrt{{1 \over d^2} + {4 \over L^2}}$ that 
of the TE$_{012}$ mode.  Since $\omega_1 \simeq m_a$, Eq.~(\ref{less}) 
becomes 
\begin{eqnarray} 
P_w &=& 4.8~{\rm kW} \left({\nu_0 \over 10~{\rm MHz}}\right) 
\left({10^{10} \over Q_w}\right)
\left({10^8 \over Q_1}\right)^2 \epsilon^2 \nonumber\\
&~&\left({T_{\rm n} \over 10~{\rm K}}\right)
\left({\rho_a \over {\rm GeV/cc}}\right)
\left({g_\gamma \over 0.36}\right)^2
\left({s/n \over 15}\right)~\ .  
\label{dispow} 
\end{eqnarray}
$\nu_0 = {\omega_0  \over 2 \pi}$ and $L$ are related by 
$L = 15 {\rm m} \left({10 {\rm MHz} \over \nu_0}\right)$. The 
surface resistance of superconductors typically varies with 
frequency $\nu_0$ as $\nu_0^{1 \over 2}$.  Since $Q_w \sim 10^{10}$ 
is achieved at GHz frequencies, we expect 
$Q_w \sim 10^{10} \left({{\rm GHz} \over \nu_0}\right)^{1 \over 2}$.
 
For definiteness, we consider a cavity with $L = 5$m, $\nu_0$ = 30 MHz, 
$Q_w = 5 \cdot 10^{10}$ and hence $P_w$ = 2.9 kW.  The electrical power 
consumption of the refrigerator is then of order 0.8 MW.  The 
microwave power circulating through the cavity is $P_0$ = 1.45 MW.  As 
was already mentionned, up to 99\% of $P_0$ may be recycled.  The energy 
stored in the cavity is $E_0 = {P_0 \over \omega_0} Q_0$ = 0.77 MJ.  To 
achieve the high value of $Q_w$ envisaged, the electric field of the 
input TEM mode must be less than approximately 20 MV/m \cite{Padamsee}.  
Equivalently, its energy density must be less than 1.77 mJ/cm$^3$.  Hence 
the cavity volume $V$ must be larger than 435 m$^3$ or its cross-sectional 
area $A = V/L$ larger than 87 m$^2$.  To achieve such a large volume the 
actual cavity must consist of an array of subcavities coupled together, 
each with cross-sectional area similar to that shown in Fig. 1.

The present proposal was motivated in large part by the desirability of a
practical method to extend the search for axion dark matter to larger axion 
masses than has hitherto been achieved with the well established method using 
a cavity permeated by a static magnetic field.  Whether the proposed method
achieves that is not clear.  We found that the energy costs are manageable, 
but the cavity has to be large.  Furthermore the following challenges must 
be overcome.  First, it has to be shown that non-linearities in the cavity 
response do not pump an excessive fraction of the input power into the 
signal mode.  The allowed fraction is of order $10^{-23}$. Second, a 
suitable tuning mechanism has to be devised.  It must be able to control 
the frequency with a precision of $Q_1^{-1} \sim 10^{-8}$.  Third, the 
signal of a few times $10^{-20}$ W at a frequency of a few GHz has to be 
filtered out of a few kW of input power at a frequency of order 30 MHz.

I am grateful to Aaron Chou, Guido Mueller, Neil Sullivan, 
David Tanner, Karl van Bibber, Leslie Rosenberg and Frtiz Caspers 
for useful comments.  This work was supported in part by the U.S. 
Department of Energy under contract DE-FG02-97ER41029.

\end{document}